# MICRO SPITBOL

Robert B. K. Dewar[a], Martin Charles Golumbic[b], and Clinton F. Goss[c]

[a] AdaCore, 104 Fifth Avenue, 15th floor, New York, NY 10011. Email: dewar@adacore.com
[b] The Caesarea Rothschild Institute, University of Haifa, Mt. Carmel, Haifa, 31905. Email: golumbic@cs.haifa.ac.il
[c] Westport, CT 06880. Email: clint@goss.com

**ABSTRACT**

A compact version of MACRO SPITBOL, a compiler/ interpreter for a variant of SNOBOL4, has been developed for use on microcomputer systems. The techniques for producing an implementation are largely automatic in order to preserve the integrity and portability of the SPITBOL system. These techniques are discussed along with a description of an initial implementation on a 65K byte minicomputer. An interesting theoretical problem which arises when using procedures which compact the interpretive object code is also analyzed.

### Preface

This is a revised edition of:

Robert B. K. Dewar, Martin Charles Golumbic, and Clinton F. Goss. *MICRO SPITBOL*. Computer Science Department Technical Report No. 11, Courant Institute of Mathematical Sciences, New York University, October 1979, 17 pages.

Stylistic and layout changes as well as typographical corrections have been made, but the content is substantially identical to the initial publication. The text for this edition was provided courtesy of The Internet Archive.

### Preamble of the Initial Publication

This material is based upon work supported in part by the National Science Foundation under Grant No. MCS78-03820.

Any opinions, findings, and conclusions or recommendations expressed in this publication are those of the authors and do not necessarily reflect the views of the National Science Foundation.

### Introduction

MACRO SPITBOL (1) is a compiler/ interpreter for a variant of SNOBOL4 (3), SPITBOL (4), which has been implemented on a variety of large computers. MICRO SPITBOL is an adaptation of the system for use on micro and minicomputers which compiles a language identical to MACRO SPITBOL with the exclusion of real arithmetic. The goal was to preserve the structure and machine independence of the system while allowing for the added constraints imposed by small computers, particularly the severe limitations on memory and address size. We also intended that the process for implementing MICRO SPITBOL on a minicomputer be largely automated. In this way we preserve the integrity of the well-tested MACRO SPITBOL source code while allowing for easy updating to new versions.

These goals were attained by encoding the MACRO SPITBOL source code, written in the MINIMAL assembly language (5), into a compact microcomputer assembly language, MICRAL. In a MACRO SPITBOL implementation, the MINIMAL source code is translated into the target machine's assembly language and directly executed; The target MICRO SPITBOL machine emulates a standard virtual microcomputer (MICRAC) which executes MICRAL code. This interpretive approach preserves the portability of MINIMAL programs while dealing with the problems and constraints imposed on MICRO SPITBOL.

MICRO SPITBOL is implemented on the Incoterm SPD 20/40 (6), a 65K byte minicomputer, and versions are in preparation for microcomputers based on the IMS 8080 and M6800 processors. The Incoterm implementation has been exercised with an extensive test package and appears very sound. It provides about 2/5 of memory for the user heap in which 400 to 500 SPITBOL statements may be compiled. The current version executes an average of 20 SPITBOL statements per second; This is 1/140 the execution speed of the CDC 6600 version.

## MICRAL - Micro Computer Assembly Language

MICRAL (2) is an encoding of MINIMAL suitable for direct translation to an interpretive form to run on a mini or microcomputer system with a 16 bit addressing structure and 8 bit bytes. The layout of a MICRAL program and the environment in which it runs are similar to that of a MINIMAL program. However, MICRAL source code is largely non-symbolic; Most of the work in converting MICRAL to its interpretive form involves address resolution, a feature available on even the most primitive assemblers.

Unlike MINIMAL, MICRAL assumes no minimum hardware configuration since its machine code is interpretive. In fact, the object code for a given MICRAL program does not vary among machines with the exception of absolute addresses. Thus the sophistication of the target machine's instruction set determines only the speed at which the MICRAL object code is interpreted and not the size of the code, as in MINIMAL.

The general format of MICRAL source code is typical of many assembly languages. The opcode is one of 80 mnemonic instructions which is translated into a single byte (x'00' to x'4F'). The operand field is a (possibly empty) series of items separated by commas. Each item is one of the following:

1) A two or four digit hexadecimal number which is assembled into one or two object bytes;

2) A five character label appearing in the MICRAL program or a system subroutine. This is eventually translated into the two byte absolute address of the label;

3) A five character label proceeded by a '+' or '–'. This form of an address reference is used to generate a special single byte address which will be discussed later.

To represent MINIMAL operands in MICRAL, the byte following the opcode is used to specify the format of the first two operands. This byte, the operand header byte, contains information about the first operand in its four low order bits and information about the second operand in its four high order bits. In frequent cases (references to the contents of the six virtual registers), the MINIMAL operand is completely specified in the header byte.

For example, the MINIMAL instruction < MOV WA,-(XS) > is encoded in MICRAL as < MOV 90 > since x'0' is the four bit code for the contents of the WA register and x'9' indicates the contents of the word indexed by (XS) with a pre-decrement on XS. If additional fields are required to fully specify a MINIMAL operand, they directly follow the operand header byte.

The fields for the first two operands may be followed by a field for the possible third operand. For example, the MINIMAL < BEQ WA,=X'00FF',LABEL > would be encoded as < BEQ B0,00FF,LABEL > since x'B' indicates a literal value for the second MINIMAL operand.

In accordance with the space efficient philosophy of MICRAL there are several modifications we apply to the language as we have described it.

First we restrict all literal values and offsets to be in the range 0 to x'7FFF'. This allows literals in the range 0 to x'7F' to be expressed as a single byte whose value is x'80'+literal.

In a similar style, we convert local memory references to a single byte by requiring that all labels in the MICRAL program and system routines take on a value less than x'8000'. We express a jump address, ADR, which is in the range (L − x'40') to (L + x'3F') (where L is the current location) as a single byte containing (L + x'C0' − ADR). These short branch instructions are recognized during the translation from MINIMAL to MICRAL and are indicated by substituting '–LABEL' or '+LABEL' for 'LABEL' whenever LABEL is in range.

The final and most space efficient optimization employed is the use of macros. MICRAL macros utilize the 176 unused opcodes x'50' to x'FF' for representing sequences of bytes in the MICRAL object code. These sequences of bytes may be a single MICRAL instruction, part of an instruction, or several instructions. Note that macros never include bytes corresponding to the one byte offset form of jump addresses, so the question of where to count the offset from does not arise. Also, a macro opcode may never be substituted for a sequence of bytes which have an embedded label in the source code. This eliminates jumping into the 'middle' of a macro. Also, it is not possible to embed a macro in another macro, thus eliminating the necessity for environment stacking when interpreting the macros.

Macros may be interpreted by the MICRAC emulator in one of two ways. A table of the MICRAL bytes corresponding to each macro may be referenced when a macro is encountered. The emulator interprets the bytes obtained from the macro table as if they had appeared in the MICRAL program. The number of bytes for each macro is also maintained within the macro table and, when the macro bytes are exhausted, interpretation returns to the main byte string.

Alternately, some or all of the macros may be directly hand encoded into the target machine language. This latter approach requires more space, but increases execution speed.

**The PD/FMS Implementation**

The Incoterm SPD 20/40 is a byte oriented single address minicomputer with indexing and multi-level indirect addressing. It supports up to 65K bytes of magnetic core memory with a cycle time of 1.6 microseconds. MICRO SPITBOL is implemented under the Program Development/File Management System (PD/FMS) (7) which provides for task control and device handling via a package of relocatable routines.

Minimum hardware for running MICRO SPITBOL consists of the central processor with 65K bytes of memory, a CRT/keyboard unit, and a 10 megabyte cartridge disk drive. Additional supported hardware includes 7 CRT/keyboard units, a printer, and 3 cartridge disk drives.

The details of the PD/FMS implementation were as follows:

The MACRO SPITBOL source code, written in MINIMAL, was translated to MICRAL source code by MICTRAN, a program coded in SPITBOL. It is a variation of a SPITBOL translator used for generating native assembly code from MINIMAL. It recognizes the short form of literals and branch instructions as described earlier. The program was run on a CDC 6600 and processed the 10,000 MINIMAL instructions in 20 CPU seconds. Figure 1 is an example of the code produced by MICTRAN. The MICRAL source code, along with the error text file, was then exported to the Incoterm system where the remainder of the implementation was done.

The next stage was to determine the sequences of macros to employ in translating the MICRAL source code to SPD assembly code (8). We first examined the problem theoretically to determine whether there is an algorithm for giving optimal or near optimal results. The problem is interesting since, although the computation may be performed in polynomial time, the complexity of the optimal algorithm is a polynomial whose degree depends on the number of macros chosen: finding the $\upsilon$ macros which minimize the length of $\eta$ bytes of object code is $O((\eta \ell)^{\upsilon+2} / (\upsilon-2)!)$ where $\ell$ is the maximum length of a macro. Another

```
       EXIXR                          EXIXR  RTN
             MOV  94                         MOV  XR,-(XS)     STACK RESULT
       EXITS                          EXITS  RTH
             LCW  04                         LCH  XR           LOAD CODE WORD
             MOV  37                         MOV  (XR),XL      LOAD ENTRY ADR
             BRI  03                         BRI  XL           EXECUTE NEXT CODE
       EXNAM                          EXNAF  RTN
             MOV  93                         MOV  XL,-(XS)     STACK NAME BASE
             MOV  90                         MOV  WA,-(XS)     STACK NAME OFSET
             BRN  EXITS                      BRN  EXITS        DO NEXT CODE WORD
       EXNUL                          EXNUL  RTN
             MOV  9B,NULLS                   MOV  =NULLS,-(XS) STACK NULL VALUE
             BRN  EXITS                      BRN  EXITS        DO NEXT CODE WORD
       EXSID                          EXSID  RTN
             MOV  0A,25                      MOV  CURID,WA     LOAD CURRENT ID
             BNE  B0,7FFF,+EXSI1              BNE  WA,=CFP$M,EXSI1  JUMP NO OVFL
             ZER  00                         ZER  WA           RESET FOR WRAP
       EXSI1 ICV  00                  EXSI1  ICV  WA           BUMP ID VALUE
             MOV  A0,25                      MOV  WA,CURID     STORE FOR NEXT
             MOV  E0,82                      MOV  WA,IDVAL(XR) STORE ID VALUE
             BRN  EXIXR                      BRN  EXIXR        EXIT WITH RESULT
```

**Figure 1**.  Example of MINIMAL code (right) translated to
MICRAL (left) by MICTRAN

approach which produces near optimal results and does allow embedding of macros requires $O(\upsilon(\eta+\ell+\eta\ell\log_2\eta))$ time. The analysis of this problem is presented in the appendix.

In practice, this theoretical complexity is troublesome since version 3.3B of MICRO SPITBOL requires 23,110 bytes of MICRAL object code before macro substitution. Also, implementation of the algorithms analyzed in the appendix is difficult due to the numerous constraints on the allowable macros. Since macros may not contain short branches or embedded labels, choosing macros must be done from the source code and not the resulting byte string. Also, since additional short branches become available as macros are introduced, the computation of the length of the object code after a macro substitution is dependent on many factors.

Due to the desire for a speedy implementation, it was decided to restrict macros to a single instruction or part of an instruction and no use of hand encodings or short branches were made. The MICRAL source code was sorted lexicographically with the opcode field as the first key and the operand field as the second key. The sorted list was then scanned by hand to count the number of each operational MICRAL instruction or part instruction. The number of bytes saved by converting each sequence to a macro was then computed for each such instruction and multiplied by the number of occurrences. The 176 byte sequences which produced the highest savings were chosen as macros.

The MICRAL source code and the chosen macros were then processed by MICASM, which translates MICRAL to SPD assembly source code (see Figures 2 and 3). This program is coded in SPD assembly language and processes 65 lines per second. The resulting assembly language file was then assembled into a relocatable module. The size of that module with macro substitutions is 17,920 bytes, an improvement of 5,190 bytes over the corresponding module without macro substitutions.

To interpret the MICRAL object code, another module was included in the absolute assembly. This module, known as the MICRAC emulator, consists of 3,020 lines of code and takes 2,230 bytes of memory. Since its size is small compared to the MICRAL object code, the emulator is one area of the system in which speed considerations took priority over space.

Finally, a set of modules consisting of PD/FMS routines for task control, timer control, and screen, disk, and printer

```
        EXIXR   EQU   $              EXIXR
                HEX   3294                   MOV   94           STACK RESULT
        EXITS   EQU   $              EXITS
                HEX   2A04                   LCW   04           LOAD NEXT CODE WORD
                HEX   3237                   MOV   37           LOAD ENTRY ADDRESS
                HEX   0B03                   BRI   03           EXECUTE NEXT CODE
        EXNAM   EQU   $              EXNAM
                HEX   3293                   MOV   93           STACK NAME BASE
                HEX   3290                   MOV   90           STACK NAME OFSET
                HEX   030C                   BRN   EXITS        DO NEXT CODE WORD
                ADDR  EXITS
        EXNUL   EQU   $              EXNUL
                HEX   329B                   MOV   9B,NULLS     STACK NULL VALUE
                ADDR  NULLS
                HEX   030C                   BRN   EXITS        DO NEXT CODE vlORD
                ADDR  EXITS
        EXSIO   EOU   $              EXSIO
                HEX   320A25                 MOV   0A,25        LOAD CLIRR ENT 10
                HEX   09B07FFF               BNE   B0,7FFF,+EXSI1
                ADDR  EXSI1
                HEX   4400                   ZER   00           RESET FOR WRAP
        EXSI1   HEX   1C00           EXSI1   ICV   00           BUMP IO VALUE
                HEX   32A025                 MOV   A0,25        STORE FOR NEXT TIME
                HEX   32E082                 MOV   E0,82        STORE ID VALUE
                HEX   03                     BRN   EXIXR        EXIT WITH RESULT
                ADDR  EXIXR
```

**Figure 2**. Example of translation from MICRAL to SPD assembly code with no macro substitution.

handling were included in the absolute assembly. These routines are accessed by SPITBOL through a set of system routines which are called directly by the MICRAL code. The system routines were coded in SPD assembly language and are part of the mainline MICRO SPITBOL file (the absolute assembly). In addition to the system routines, this file contains the macro sequences and allocates space for the relocatable modules and the SPITBOL heap. It is 5140 lines of code and occupies the full 65K bytes, 28k bytes of which are the user heap.

The PD/FMS implementation has been exercised on a variety of test programs and appears very stable. Performance evaluation was done with a package of 24 programs which have an average length of 70 statements. The test set was run under version 3.3B of MACRO SPITBOL on the CDC 6600 for comparison. MICRO SPITBOL compiled at an average rate of 1.3 statements per second and executed at 19.7 statements per second. The corresponding values for MACRO SPITBOL were 186.9 statements per second for compilation and 2,829 statements per second for execution.

## Conclusion

The implementation of MICRO SPITBOL was broken down into two major stages.

First, the design and implementation of programs for converting from MINIMAL to the interpretive MICRAL code, choosing appropriate macros, and converting from MICRAL to assembly language, required about three man-months work.

The second stage entailed work specific to the PD/FMS implementation. This included coding the MICRAC emulator and system routines in SPD assembly language which, for the PD/FMS implementation, took about 1 man-month work. For subsequent implementations it is only this latter stage which is needed since the MICRAL object code and macros are fixed for a given version of MICRO SPITBOL and only minor changes in MICASM are necessary to accommodate various native assembly language formats.

Also, since the process of translation for MICRAL is now automated, any program coded in MINIMAL may now be implemented by this scheme.

The PD/FMS version currently allows 400 to 500

```
        EXIXR   EQU   $             EXIXR
                HEX   6A ***               MOV   94          STACK RESULT
        EXITS   EQU   $             EXITS
                HEX   2A04                 LCW   04          LOAD NEXT CODE WORD
                HEX   89 ***               MOV   37          LOAD ENTRY ADDRESS
                HEX   0B03                 BRI   03          EXECUTE NEXT CODE
        EXNAM   EQU   $             EXNAM
                HEX   6B ***               MOV   93          STACK NAME BASE
                HEX   6D ***               MOV   90          STACK OFSET
                HEX   E1 ***               BRN   EXITS       DO NEXT CODE WORD
        EXNUL   EQU   $             EXNUL
                HEX   68 ***               MOV   9B,NULLS    STACK NULL VALUE
                ADDR  NULLS
                HEX   E1 ***               BRN   EXITS       DO NEXT CODE WORD
        EXSID   EQU   $             EXSID
                HEX   9A ***               MOV   0A,25       LOAD CURRENT ID
                HEX   25
                HEX   EC ***               BNE   B0,7FFFF,+EXSI1
                HEX   7FFF
                ADDR  EXSI1
                HEX   4400                 ZER   00          RESET FOR WRAP
        EXSI1   HEX   1C00          EXSI1  ICV   00          BUMP ID VALUE
                HEX   67 ***               MOV   A0,25       STORE FOP NEXT TIME
                HEX   25
                HEX   5D ***               MOV   E0,82       STORE ID VALUE
                HEX   E0 ***               BRN   EXIXR       EXIT WITH RESULT
```

**Figure 3**. Example of translation from MICRAL to SPD assembly code with macro substitutions.  <\*\*\*> indicates the use of a macro.

'typical' SPITBOL statements to be compiled. Again, since the MICRAL object code does not vary among implementations, this figure should remain fairly constant for microcomputers with 65K bytes of memory.

The cost of the extra level of interpretation necessary to emulate the MICRAC machine is evident in the two order drop in execution speed from MACRO SPITBOL on the CDC 6600 to the PD/FMS implementation. However, the Incoterm instruction set is rudimentary compared to other target microcomputers, so an improvement in execution speed may be expected on other processors. Extra speed may also be gained by utilizing some of the unused design features of MICRAL such as hand encodings for macros. Also, a better selection of substitution macros will have a positive effect on both processing speed and space. Better macros may be selected by implementing one of the algorithms suggested in the appendix, with the possible addition of some efficient heuristics.

Due to the decreasing cost of memory, the space constraints experienced in this implementation will probably disappear. Memory may become so inexpensive that a direct translation of the MINIMAL code to the target machine code would be preferable for small MINIMAL programs. However, for large MINIMAL programs, or on processors with a low level instruction set, the techniques used in the implementation of MICRO SPITBOL may be more suitable. Since these systems are typically single-user, the availability of very high level languages outweighs the loss of processing speed for many applications.

## Appendix

We now analyze the problem of finding a set of macro substitutions which minimizes the space required for the object code and macro table. We ignore the complications which arise when the short form of branches are used in the object code byte string since we have not implemented this aspect of MICRAL.

Let $B = \langle b_1, \ldots, b_\eta \rangle$ be a sequence of bytes. The length of B is denoted $|B| = \eta$. A subsequence of B, $\langle b_i, \ldots, b_j \rangle$, is denoted $B(i:j)$.

A macro set is a set $M = \{m_1, \ldots, m_\upsilon\}$ where each macro $m_i$ is a sequence of bytes. The size of the macro table corresponding to M is given by

$$\|M\| = \sum_{i=1}^{\upsilon} |m_i|$$

Since macros may not include a labeled byte other than the first byte, macros tend to be short compared to $|B|$. To aid in the analysis, we restrict the length of macros, $|m_i| < \ell$.

A macro substitution is an operation on a byte sequence in which all occurrences of a given subsequence of bytes are replaced by a single byte. In case of overlapping occurrences of the subsequence, the leftmost occurrence is replaced. The byte sequence resulting from the substitution of all macros in a macro set, M, on a byte sequence, B, is denoted $B(M)$. The goal is to find, for a given byte sequence, B, a macro set, M, which minimizes the length function

$$L(B,M) = |B(M)| + \|M\|$$

We shall analyze the worst case involved in a complete search strategy. For $\upsilon=1$, the problem reduces to finding the best single macro for a byte sequence. Algorithm A performs this function using a table, FREQ, which maps subsequences of B onto a count of their occurrences in B.

*Algorithm A:*

1) For k from 1 to $\ell-1$ perform steps 2 and 3.

2) For each subsequence of B, $B(i:i+k)$, increment $FREQ(B(i:i+k))$.

3) Set $m(k)$ to a subsequence such that $FREQ(m(k))$ is maximal. Then set:

   $L(B,\{m(k)\}) = |B| - (|m(k)|-1)*(FREQ(m(k))-1) + 1$

4) The best macro is the $m(i)$ which minimizes $L(B,\{m(k)\})$ in step 3,

5) Substitute the macro $m(i)$ in B.

Steps 2 and 3 are iterated over $O(\ell)$ times. Step 2 requires $O(\eta)$ time to scan the subsequences and the map FREQ, if implemented as a balanced binary search tree, requires $O(\log_2 \eta)$ to maintain. Step 3 is linear if, during step 2, a pointer to the macro with the greatest FREQ is kept. Step 4 may also be done in this fashion in linear time. Performing step 5 may be done by a modification of the deterministic finite automaton pattern matching algorithm presented in Morris and Pratt (9) and expanded in Aho, Hopcroft, and Ullman (10). This algorithm requires $O(|B| + |m(i)|) = O(\eta + \ell)$ time. Note that we perform step 5 by finding the leftmost occurrence of $m(i)$, substitute for it, and continue to the right.

Therefore, Algorithm A runs in $O(\eta + \ell + \eta \ell \log_2 \eta)$ time.

For $\upsilon > 1$, one would expect that an iterative application of Algorithm A would be effective. However, an optimal choice for the first macro does not necessarily lead to an optimal macro set. For example, if

B=<j,a,b,c,d,e,f,m,r,h,a,b,c,d,e,g,k,c,d,e,f,n,s,h,a,b,c,p>

a choice of $m_1$ = <a,b,c,d,e> yields $L(B,\{m_1\})$ = 25 which is optimal for $\upsilon$=l. However, any choice for $m_2$ will force $L(B,\{m_1,m_2\}) > 25$. If $m_1$ = <c,d,e,f> and $m_2$ = <h,a,b,c>, then $L(B,\{m_1\})$ = 26 and $L(B,\{m_1,m_2\})$ = 24.

Furthermore, our algorithm for pattern matching may not produce optimal results in cases where a macro overlaps itself. Substituting for the second occurrence in an overlapping pair may lead to a better choice for other macros. This forces the recognition of each occurrence of a macro and treatment of overlapping occurrences as separate cases.

We now characterize the problem of finding an optimal macro set by a graph theoretic representation.

A weighted graph G = (V, E, W) consists of a binary relation E over a finite set V of vertices, and an integer valued mapping W defined over V. G is an interval graph if its vertices can be put into 1–1 correspondence with a set of intervals on a line so that two vertices are adjacent if and only if their corresponding intervals intersect. If B is a sequence of bytes, G(B), the <u>weighted</u> <u>interval</u> <u>graph</u> corresponding to B, is defined as follows: Each vertex in V represents a substring of B which may be used as a macro; Vertices $v_i$ and $v_j$ are connected by an edge in E iff the occurrences of the substrings represented by $v_i$ and $v_j$ overlap in B. If V is a vertex representing an occurrence of the substring m, then $W(v) = |m| - 1$.

If S is a subset of vertices from V, the subgraph of G induced by S, denoted $G_S$, consists of all vertices of S together with all edges from E which connect those vertices.

To construct G(B), the $O(\eta\ell)$ possible macros in B are scanned and one vertex is added to V for each. Whenever a vertex is added to V, edges are added to E for each overlapping macro pair. Since adding edges takes $O(\ell^3)$ time with efficient data structures for the graph, G(B) is constructed in $O(\eta\ell^4)$ time.

To find the macro set, M, which minimizes L(B,M), we apply the following algorithm:

*Algorithm B:*
1) Calculate G(B).
2) Partition the vertices of G(B) such that each partition contains vertices which represent occurrences of identical substrings in B.
3) Let C be the set of all possible combinations of $\upsilon$ partitions. For each set of combinations in C perform step 4.
4) Let S be the vertices in the $\upsilon$ partitions. Let $M_S$ be the macro set consisting of the $\upsilon$ distinct byte sequences corresponding to the vertices of S. Occurrences of the macros in $M_S$ are chosen in B such that no occurrences overlap. The problem of choosing the best set of occurrences is analogous to finding the maximum weighted independent set of vertices, $I_S$, in $G_S$. Then

$$L(B, M_S) = |B| - \sum_{v \in I_S} W(v) + \sum_{s \in S} W(s) .$$

5) The macro set, M, which minimizes L(B,M) is then determined by choosing an $M_S$ which minimizes $L(B, M_S)$ in step 4.

Gilmore and Hoffman (11) characterized interval graphs by showing them to be a subclass of chordal graphs. This class of graphs has the property that, for every simple cycle $[v_1, v_2, \ldots, v_n, v_1]$ (n>3), there is an edge $(v_i, v_j)$ in E such that $v_i$ and $v_j$ are in the cycle but the edge is not.

Gavril (12) gives an $O(|V|^2)$ algorithm for finding the maximum weighted independent set of a chordal graph. The algorithm is also presented in Golumbic (13) with suitable data structures. See also Booth and Leuker (14).

Thus, step 3 may be performed in $O((\upsilon\eta)^2)$ time (note that the number of vertices in any induced subgraph is bounded by $\upsilon\eta$). Step 3 must be performed at most $\binom{\eta\ell}{\upsilon}$ times. Hence, the worst case time complexity of algorithm B is

$$O((\eta\ell^4) + (\upsilon\eta)^2 \binom{\eta\ell}{\upsilon}) ,$$

which is, in terms of the length of the input string, a polynomial whose degree depends on the constant $\upsilon$. In the example of this paper, $\eta$ is 23,000, $\upsilon$ is 176, and a reasonable $\ell$ would be 20. With values in this range, the above complexity approximates and is bounded by

$$O((\eta\ell)^{\upsilon+2}) / (\upsilon-2)!) .$$

Since the time complexity of Algorithm B is high for this application, an effective algorithm for finding a near optimal solution is needed. One such heuristic approach would', be an iterative application of Algorithm A whose time complexity is

$$O(\upsilon(\eta+\ell+\eta\ell\log_2\eta)) .$$

Such an approach may potentially produce better results than Algorithm B since it allows the embedding of macros within other macros, a feature prohibited in Algorithm B. Although, one could theoretically include this embedding feature in a method similar to Algorithm B by using the weighted overlap graph model as described in Gavril (15) and Golumbic (13).